\documentclass[aps,prd,showpacs,twocolumn,amsfonts]{revtex4-1}
\usepackage{epsfig}
\usepackage{graphics}
\usepackage{amsmath}
\usepackage{amssymb}
\usepackage{bm}

\newcommand{\be}{\begin{equation}}
\newcommand{\ee}{\end{equation}}
\newcommand{\bea}{\begin{eqnarray}}
\newcommand{\eea}{\end{eqnarray}} 

\newcommand{\Eq}[1]{Eq.~(\ref{#1})}
\newcommand{\Eqs}[1]{Eqs.~(\ref{#1})}

\newcommand{\half}{\frac{\scriptstyle 1}{\scriptstyle 2}}

\newcommand{\ket}[1]{| #1 \rangle}

\begin{document}

\hyphenation{}

\author{Jacob D. Bekenstein}
\affiliation{Racah Institute of Physics, Hebrew University of
Jerusalem, Jerusalem 91904, Israel\\}\date{\today}
\title{Is a tabletop search for Planck scale signals feasible?}
\begin{abstract}
Quantum gravity theory is  untested experimentally.  Could it be tested with tabletop experiments?  While the common feeling is pessimistic, a detailed inquiry shows it possible to sidestep the onerous requirement of localization of a probe on Planck length scale.  I suggest a tabletop experiment which, given state of the art ultrahigh vacuum and cryogenic technology, could already be sensitive enough to detect Planck scale signals.  The experiment combines a single photon's degree of freedom with one of a macroscopic probe to test Wheeler's conception of ``quantum foam'', the assertion that on length scales of the order Planck's, spacetime is no longer a smooth manifold.  The scheme makes few assumptions beyond energy and momentum conservations, and is not based on a specific quantum gravity scheme.
\end{abstract}
\maketitle

\section{Introduction} \label{sec:intro}

The theory of quantum gravity is a salient \emph{desideratum} of contemporary physics.  But the proposed quantum gravity schemes differ considerably from one another and are remote from experimental confrontation.   Given this situation one would like to have tests of the basic ideas which inspire the search for the quantum gravity-to-be.  One of these is Wheeler's proposal of ``quantum foam''~\cite{Wheeler}: on scales below that of Planck, spacetime is no longer a smooth manifold, but rather a frothy and tumultuous landscape.  How can this idea be tested without assuming too much about quantum gravity theory?  

During the last decade a sizable fraction of investigators adopted the view that the palpable world is confined to a four dimensional brane in higher dimensional spacetime, with the additional dimensions not necessarily microscopic~\cite{RS}.  This implies that the fundamental 4-D Planck length, $\ell_P$, is really much larger than the quantity $\ell_P=(\hbar G/c^3)^{1/2}=1.616\times 10^{-35}\,{\rm m}$, where $G$ is the measured Newton constant.  Can we decide experimentally between this scheme and the traditional one?  Can we decide experimentally for or against the proposal that gravity, being a kind of thermodynamics of spacetime~\cite{Jacobson}, is in no need of quantization?

Hopes have been expressed that the Planck scale can be probed through its influence on the inflationary stage of the universe soon after the big bang~\cite{Rinaldi}, through its modification of the energy-momentum dispersion relation or the uncertainty principle as put in evidence by ultra-high energy astrophysical processes~\cite{Matt}, through the residual noise in LIGO or other gravitational waves interferometers~\cite{Ng}, and through the possible formation of microscopic black holes in the LHC particle collider~\cite{Park}.  Thus far none of these approaches has yielded unambiguous evidence of Planck scale physics.

Could one use laboratory experiments to probe Planck scale physics?  Some suggestions in this directions focus on experimental consequences of deformations of the familiar uncertainty relation which are expected by many to be important near Planck scales~\cite{Ali}.  If one is rather interested on direct evidence for quantum foam, it goes almost without saying that detection of such on Planck scale is unfeasible with an elementary particle as probe.  If we tried to localize the particle to the required scale, the uncertainty principle would require that we give the particle a momentum of at least $\hbar/ \ell_P$, which for elementary particle masses corresponds to an energy of at least $10^{19}$ GeV; this is many orders of magnitude beyond what foreseeable particle accelerators will afford. 

Could one instead ``see'' the quantum foam using a macroscopic probe (mass $M$), say, by observing the effect of moving it a distance of order $\ell_P$?  Again the answer is negative if the experiment involves localizing the probe (more correctly, its center  of mass, c.m. henceforth) to better than a Planck length, so that we can be sure that the probe, as a whole,  has moved only by a distance of that order.  According to the uncertainty principle, such localization would introduce an uncertainty $\Delta p>\hbar/\ell_P$ in the probe's momentum.   Thus to engender a translation \emph{under control} we would have to give the probe at least $\hbar/\ell_P$ of momentum, which would change its velocity by at least $\hbar/(M \ell_P)$.  During a time interval $\tau$ the probe would move an uncontrolled distance of at least $\tau\hbar/(M \ell_P)$.   This would remain smaller than $\ell_P$ only if $\tau< (\ell_P/c)(M/m_P)$, where $m_P=(c\hbar/G)^{1/2}\approx 2.177\times 10^{-8}\,$Kg in the no extra dimensions scenario.  Even for $M\sim 10^3\,$Kg, $\tau$ would have to be shorter than $10^{-32}\,{\rm s}$.  But switching a device on and off that fast is beyond foreseeable technology.

However, one could succeed in the envisaged task using a macroscopic probe if its operation did not depend on localizing it with any great accuracy, for then the preceding argument would be rendered irrelevant. But how could one be sure that the probe moves a distance of order $\ell_P$ without first localizing it?  By relying on conservation of momentum. 

In what follows we propose the idea for a table-top experiment which, depending on the outcome, may confirm the radical texture of sub-Planckian spacetime, and decide whether the Planck scale is very small or merely microscopic.  The idea, in brief, is to use a single optical photon which traverses a dielectric block to engender a translation of the block which can be arranged to be of order the Planck scale.  The translation \emph{does not} hinge on giving a permanent impulse to the block.  Certification that the tiny translation actually occurred is to be had from detecting the photon after transit through the block and relying on momentum conservation.  But, as discussed in Sec.~\ref{transits}, translation by a distance of order $\ell_P$ is expected to be impeded with some probability.  Thus if in a series of like experimental runs the frequency with which the photon is found to get through the block falls short of expectations (from the block's classical transmission coefficient), this may signal that spacetime is ``rough'' at the relevant scale.  The scale at which spacetime ceases to be smooth could thus be experimentally determined.   
  
\section{The ideal experiment}\label{ideal}

\begin{figure}
\includegraphics[width=3.0in]{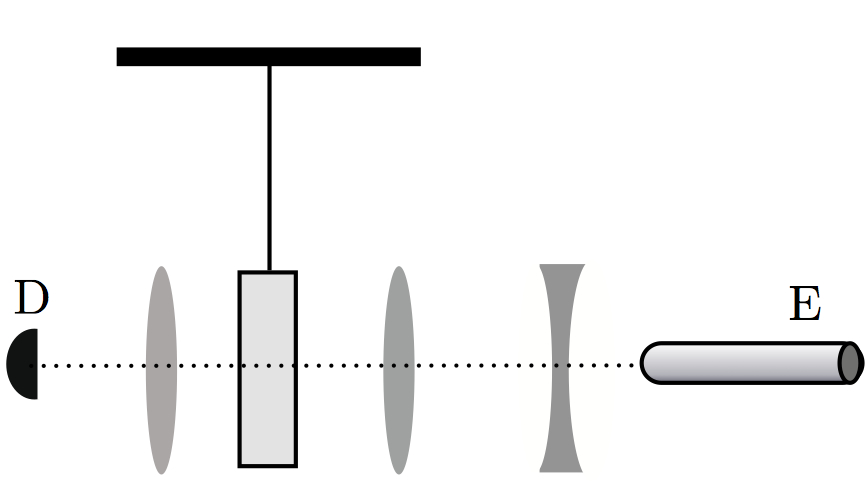}
\caption{\label{Fig1}Showing the suspended block and the photon's path (dotted) starting from the single-photon emitter E, passing through a divergent$\,+\,$convergent lens system (to spread the pulse) and, after transiting the block, passing through a second convergent lens that focuses the pulse onto the single-photon detector D.}
\end{figure}

The macroscopic probe shall be a rectangular dielectric block of dimensions $L_1\times L_2\times L_2$ and mass $M$ that may be crystalline or amorphous.  We require it to be highly transparent to optical electromagnetic waves.  The dielectric is supposed optically isotropic, e.g. if a mono crystal it should be of the cubic crystal class.   The block shall be suspended from a thin fiber so that the small translation here envisioned can be regarded as frictionless motion with negligible restoring force (Fig.~\ref{Fig1}).  The fiber must be affixed to the center of the block's upper face.

A photon of vacuum wavelength $\lambda_0$ from a suitable single-photon emitter E  is to be directed at the block normally to one of the square faces (henceforth the ingress face, the opposite one being the egress face).  An optical system, also shown in Fig.~\ref{Fig1}, should shape the pulse so that it illuminates almost the whole block (as opposed to just a narrow tube through it).  A second such system, placed after the block, is used to focus the exiting pulse onto a suitable single-photon detector D.   In vacuum this photon carries momentum $p_0=h/\lambda_0=\hbar \omega/c$, where $\omega$ is the corresponding angular frequency.  What is the momentum of the photon inside the block, supposing that it was not reflected at the ingress face?  

Faraday's equation ${\bm\nabla}\times{\bm E}+c^{-1}\partial_t {\bm B}=0$, which is valid in matter as well as in vacuum, imposes on the electric field ${\bm E}$ and magnetic induction ${\bm B}$ of a plane wave which varies as $\exp(\imath{\bm k}\cdot{\bm r}-\imath\omega t)$ the relation ${\bm k}\times{\bm E}=(\omega/c){\bm B}$.  Of course, $k=n(\omega/c)$, where $n$ denotes the dielectric's index of refraction at the said frequency.   We recall that $n=\sqrt{\epsilon\mu}$, with $\epsilon$  the permittivity and $\mu$ the permeability at the said frequency, and that the magnetic field ${\bm H}$ is given by ${\bm H}={\bm B}/\mu$. Thus
\be
\sqrt{\frac{\epsilon}{\mu}}\frac{\bm k}{k}\times{\bm E}={\bm H}.
\label{EHrel}
\ee

Now in a dielectric the density of electromagnetic momentum is given by~\cite{LLED}
\be
\rho_p = \frac{|{\bm E}\times{\bm H}|}{4\pi c}=\sqrt{\frac{\epsilon}{\mu}}\frac{E^2}{4\pi c}\,,
\label{momdens}
\ee
whereas that of the energy is written as
\be
\rho_e=\frac{{\bm E}\cdot{\bm D}+{\bm H}\cdot{\bm B}}{8\pi}=\frac{\epsilon E^2+\mu H^2}{8\pi}=\frac{\epsilon}{4\pi}E^2\,,
\ee
where we have used \Eq{EHrel} and the additional fact that ${\bm k}$ and ${\bm E}$  are orthogonal.  Thus
\be
\frac{\rho_e}{\rho_p}=c\sqrt{\epsilon\mu}=c\,n\,.
\ee

We interpret the above to mean that for a single photon in a dielectric, the ratio of energy to momentum is $c\,n$.  Now in the dielectric's Lorentz rest frame  $\omega$ will be unchanged in the passage from vacuum to dielectric (or vice versa).  May we assume that the photon's energy is also unchanged?  Yes, since energy must be conserved, and there seems to be no mechanism for depositing a significant fraction of the photon's \emph{energy} in the block upon ingress (we neglect the extremely tiny kinetic energy temporarily acquired by the block and the very small Doppler redshift of $\omega$ at egress due to the slow motion of the block before it stops).  We conclude that in the block
\be
|{\bm p}|=\frac{h}{n\lambda_0}=\frac{\hbar\omega}{cn}\,.
\label{res}
\ee

This expression concurs with that of Abraham in the celebrated Abraham-Minkowski controversy~\cite{AM,Barnett}.
If one instead uses Minkowski's momentum density, replacing ${\bm E}\times{\bm H}$ in \Eq{momdens} by ${\bm D}\times{\bm B}$, one gets that $|{\bm p}|=n\hbar \omega/c$.   In what follows we shall rely on conservation of the momentum of the block plus that of the photon.  This is true for both Abraham and Minkowski photon momenta provided that one uses for the block the kinematic momentum in the first case, and the canonical momentum in the second~\cite{Barnett}.  Because the kinematic momentum is the simpler, we proceed by using the Abraham photon momentum, \Eq{res}.
  
Upon entering the block the photon obviously deposits in it forward momentum equal to
\be
\Delta p =\frac{\hbar\omega}{c}-\frac{\hbar\omega}{cn}=\frac{\hbar\omega}{c}\Big(1-\frac{1}{n}\Big)\,.
\label{follow}
\ee 
We take it that in the block's original Lorentz rest frame its c.m. (defined in the Appendix) has acquired this momentum. The c.m. thus moves with speed $\Delta p/M$ during the time interval $(n L_1/c)$ it takes for the photon to traverse the block.  Assuming the photon exits the block rather than being reflected back at its egress face, it must recover its old momentum $(\hbar\omega/c)$; the block must thus divest itself of the momentum $\Delta p$ it borrowed temporarily.   Hence upon egress of the photon the block's c.m. comes to rest in its original rest frame.  We may conclude that in this particular frame the c.m. has moved a total distance
\be
\Delta X_0 = L_1 \frac{\hbar \omega}{Mc^2}(n-1)
\label{deltax}
\ee
before coming to rest~\cite{Frisch}.  (There is no violation here of the uncertainty relation, as made clear in the Appendix.)  Certification that this translation really took place can be had by detecting the photon after its transit of the block.   

But why should we care about this small translation of the c.m.? After all the c.m. is not the position of any specific electron or quark.  \Eq{canonical} of the Appendix shows that the c.m. position components are canonically conjugate to the corresponding components of the block momentum vector.  This last is a key observable of the whole block, and so its canonical conjugate, the c.m. position observable, acts as a faithful proxy for the whole block's position.  An additional argument for the relevance of the c.m. will be given in Sec.~\ref{transits}.  Anyway, our arguments presume that what happens to the c.m. matters physically. 

How big is the translation?
As an example, let us consider a photon of wavelength $\lambda=445$ nm (energy $\hbar\omega= $ 2.78 eV), and a block of mass $0.15\times 10^{-3}\,$Kg.  This last contains 0.15 $\times$ Avogadro's number or $0.903\times 10^{23}$ nucleons, and so has mass-energy $Mc^2= 0.848\times 10^{23}$ GeV.  Thus $\hbar\omega/(Mc^2)= 3.31\times 10^{-32}$.  With $L_1= 10^{-3}\,$m and $n=1.6$ (relevant to high lead glass at $\lambda=445$ nm)  we find $\Delta X_0= 1.98\times 10^{-35}\, $m, quite close to the Planck length in the traditional (no large extra dimensions) scenario.  Probing the Planck scale with the present idea thus seems feasible.

Needless to say, the magnitude of the translation mentioned above will differ in a different Lorentz frame, and can even become of order $nL_1$ in a frame moving very fast with respect to the laboratory.  But when analyzed in the initial c.m. frame, the translation has a sharp beginning and end; in another frame translation has been going on before photon ingress and it continues after egress with the passage of the photon being marked, generically, by just a slight change in velocity.  Obviously the magnitude of such translation is not sharply limited.   The shortest c.m. translation, providing the best resolution of space-time foam as it were, is manifested in the block's rest frame.  

The above also clarifies why the experiment cannot employ a macroscopic light pulse instead of a single photon.  In the former case the pulse is always partially reflected back, and the resulting recoil of the block  imparts to it a constant velocity leading ultimately to unlimited translation.  (The consequences of back reflection of a single photon are studied in Sec.~\ref{transits} below).

A possible complication in the above considerations is dispersion in the block material.  As an example we look at Schott N-SF2 high lead glass~\cite{index}.  At $\lambda=445$ nm it has $n=1.67$ with $dn/d\lambda=-0.000256\, {\rm nm}^{-1}$.  We are interested in a photon pulse defined in time to better than $5\times 10^{-12}$ s, the nominal light crossing time for the block.  Thus the time-energy uncertainty relation lets us get by with a bandwidth $\Delta\omega=2\times 10^{12}\, {\rm s}^{-1}$ which corresponds to $\Delta\lambda=0.21$\,nm.  Consequently, with care in the preparation of the photon, the index of refraction will vary by only a fraction $3\times 10^{-5}$, so that dispersion is immaterial here.

We now consider perturbations.  Newtonian gravity of nearby objects is not a significant perturbation on the block.  It is true that a $1\,$Kg mass a distance of $1\,$m from the block will impart to it, over the course of photon transit, momentum of the same order as that given by the photon.  However, that gravitational field is to be regarded as a (minute) part of Earth's field, which determines the vertical direction for the fiber as well as the precise value of the gravitational acceleration $g$.  Thus we do not have to worry about that gravitational perturbation. 

Of course, the suspending fiber exerts a minute restoring force on the block; how does this affect the argument?  If the fiber's point of support is a distance $l$ above the block's c.m., the restoring force when the block is displaced from the vertical by horizontal distance $\delta x$ is $M g\delta x/l$.  Thus in the course of the photon transit time, $n L_1/c$, the restoring force deposits in the block  momentum $ Mg\,n\, L_1\,\delta x\,(l\,c)^{-1}$.  This becomes comparable to the momentum transferred to the block, \Eq{follow}, for $\delta x=l\hbar\omega (n-1)(M g n^2 L_1)^{-1}$.  (The same displacement is obtained if one reckons, in the rest frame in which the block was at rest, the distance travelled by the block during photon transit under the acceleration $g \delta x/l$,  and equates it to $\Delta X_0$).  With the parameters mentioned above, and assuming $l\sim 10^{-1}{\rm m}$, $\delta x\approx 7.1\times 10^{-15}\,{\rm m}$, a displacement merely the size of a nucleus.  

Even bigger displacements from the vertical result from thermal agitation, as discussed below in Sec.~\ref{which}.  But this does not mean that thermal displacement of the block c.m. from equilibrium submerges the Planck scale translation we are looking for.  Due to the low speed of sound in solid matter, a few times $10^3\, {\rm m\, s}^{-1}$, the restoring force of the fiber can only influence, during the photon transit time,  those parts of the block within about $10\, {\rm nm}$ from the point of block-fiber attachment.  This is a tiny fraction of the block volume in the examples considered in this paper (see also Sec.~\ref{cal}).  For practical reasons the illuminated part of the block will not comprise that boundary region.  Thus the motion of the block c.m. that we speak of does not include the effect of the force from the fiber.  Of course this last does play a part in establishing the motion of the block at any moment; such motion, however, is not considered here because we work in the block c.m.'s Lorentz rest frame at the moment of
 photon ingress.

\section{Transits and entanglement}  \label{transits}

Internal reflection of the photon complicates the description of the experiment.  How likely is it that the photon transits the block rather than being reflected backward, and how certain can one be that it traverses it once and not multiple times?  

According to Fresnel's relations~\cite{LLED},  if an electromagnetic wave of either polarization and electric amplitude $E_i$ propagating through a transparent medium with index $n_1$ is incident normally on the plane boundary between that medium and a second one with index $n_2$, the transmitted and reflected amplitudes are
\be
E_t=\frac{2 n_1}{n_1+n_2}E_i\qquad {\rm and}\qquad E_r=\frac{n_1-n_2}{n_1+n_2}E_i\,,
\label{Fre}
\ee
respectively.  We may interpret the above ratios as applying to the amplitude of a single photon.  Thus in the wake of a single passage through the block, our photon's incident amplitude gets multiplied by the factor
\be
F_0=\frac{4 n}{(1+n)^2}\, e^{\imath n\omega L_1/c}
\ee
where the shown phase is accrued over the thickness $L_1$ of the block (we of course neglect the block's velocity as compared to $c$).  Concurrently the block's c.m. is translated by the $\Delta X_0$ of \Eq{deltax}.

If instead the photon is back reflected at the would-be egress face and then at the ingress face a total of $j$ times before finally escaping through the egress face, its original amplitude gets multiplied by
\be
F_{j}=\frac{4n}{(1+n)^2}\frac{(n-1)^{2j}}{(1+n)^{2j}}\,e^{\imath (2j+1) n\omega L_1/c}\,.
\ee
because in addition to entering and leaving the block, it has undergone $2j$ reflections at boundaries from refraction index $n$ to index $n=1$, and has traveled the length $L_1$ a total of $2j+1$ times.  Concurrently the block's c.m. is translated by
\be
\Delta X_j = L_1 \frac{\hbar \omega}{Mc^2}(n-1+2jn)\,,
\label{deltax1}
\ee  
because every time the photon is reflected off the egress face, it imparts to the block an additional forward momentum equal to \emph{twice} the value in \Eq{res}; the said momentum causes motion of the c.m. during the time $n L_1/c$ that the photon flies backward and when the photon is reflected by the ingress face it recovers the said momentum from the block. This is repeated $j$ times.   Meanwhile the block retains the forward momentum $\Delta p$ from \Eq{follow} so long as the photon is inside it, and this last alone contributes $2j+1$ translations of size $\Delta X_0$.  The sum of the various translations is $\Delta X_j$.

If before the transit the photon's normalized state was $\ket{\gamma_i}$, after the transit the part of the state of the photon--block system  in which the photon propagates forward is, in obvious notation, 
\be
| {\psi_\leftarrow}\rangle=\sum_{j=0}^\infty \frac{4n}{(1+n)^2}\frac{(n-1)^{2j}}{(1+n)^{2j}}\,e^{\imath(2j+1) n\omega L_1/c} \,|{\gamma_i}\rangle \otimes| {\Delta X_j}\rangle\,;
\label{psi_r}
\ee
that is to say, the photon's amplitude gets entangled with the block displacement.  There is, of course, a piece $\ket{\psi_\rightarrow}$ corresponding to the photon going backwards towards its source which is not important here.

From \Eq{psi_r} the probabilities that the photon transits the block with $j$ double internal reflections are 
\be
 p_j=\frac{16n^2}{(1+n)^4}\frac{(n-1)^{4j}}{(1+n)^{4j}}\qquad j=0,1,2,\cdots
\label{getthrough}
\ee
with
\be
p_\leftarrow \equiv \sum_{j=0}^\infty p_j= \frac{2n}{n^2+1}\ .
\ee
According to Bayes' theorem the conditional probability that the photon avoided internal reflections ($j=0$) given that it transited the block is
\be
p(j=0|\leftarrow)=\frac{p_0}{p_\leftarrow}=\frac{8 n (n^2+1)}{(n+1)^4}\ .
\ee

With $n= 1.6$, as in our example, $p_{0}=0.896, p_1=0.00254,\ p_2=7.2\times 10^{-6},\ \cdots$.  In addition $p_\leftarrow = 0.899$ and $p(j=0|\leftarrow)= 0.997$.  This last is also the probability that a photon was emitted, has crossed the block without any internal reflections, and is accompanied by a shift $\Delta X_0$ of the block.   Since that probability is so close to unity we can, in this first discussion, and when the photon is detected at D, ignore the possibility of internal reflections.  The above description of the process assumes a smooth spacetime geometry.

As long as the experimental parameters are such that $\Delta X_0\gg \ell_P$, we may regard spacetime as endowed with the usual symmetries under translation, rotation and Lorentz boosts.   No impediment to the translation is expected in that case.  However, since on scales comparable to $\ell_P$ vacuum quantum fluctuations of the metric are expected to be large, we expect such fluctuations to impede translation of the c.m. by distances $\Delta X_0\approx \ell_P$. One way to visualize this is as follows.  In order of magnitude the energy density associated with order unity fluctuations of the metric on said scale is expected to be $\approx\hbar/(c\, \ell_P{}^4)$; it should have a coherence length of order Planck's.  Thus in a region with diameter of order $\ell_P$ the mass is sufficient to form a black hole of mass $\approx \hbar/(c\, \ell_P)=m_P$, the minimal black hole mass.  One may thus envisage quantum foam as a sea of virtual black holes of about Planck mass $m_P$ and Planck scale radius constantly forming and disappearing on a nearly Planck timescale.  

The block's c.m., whose translation during photon transit extends over a time long compared to Planck's, will thus frequently run into one or another such black hole.   It seems likely that this repeated interaction of the c.m. will impede its linear translation, at least for some of the photon transits.  For one, a Planck scale black hole, which is likely to be moving rapidly in a direction other than that of the c.m., is not negligibly light compared to the block's mass we have in mind, and  it must have some dynamical effect on it. The usual argument against such hindrance from Lorentz invariance of the unconfined vacuum is probably immaterial here: it is widely suspected that Lorentz symmetry is broken at the Planck scale.   We note that no black can be smaller than Planck length; this suggests that block translations much shorter than a Planck length may be immune to the hindrance just discussed.   

Incidentally, the foregoing argument is another reason for focusing on what happens to the c.m. rather than to some elementary component of the block.  We could no very well talk of a quark or electron belonging to the block colliding with a Planck sized black hole: its vastly bigger Compton length would mean the particle ``averages out'' the black hole soup and is thus insensitive to its existence.  By contrast, the Compton length associated with the c.m. of our example block is very sub-Planckian in size.

We concluded that the block's translation associated with photon transit may be impeded.  This must happen with some probability $\pi_*$.  Whenever the block's motion is impeded, the photon must be prevented from crossing the block because the associated transfer of momentum to the block and back to the photon in accordance with momentum conservation is not consistent with a block translation smaller than $\Delta X_0$.  One cannot argue that the momentum $\Delta p$ in \Eq{follow} is transferred to the black hole gas instead of being retained by the block.  The gravitational vacuum must be homogeneous on scale $L_1$ or even $\lambda$, both much larger than $\ell_P$, so that the momentum of the block plus photon must be conserved (the gravitational vacuum must have zero momentum).  Thus with probability $\pi_*$ the photon will be back reflected by the block (or absorbed).  This reflection is in addition to that required by Fresnel's formulae (or their extension to account for imperfect transparency).  

If, after accounting for the quantum efficiency of the photon detector, it is found, in multiple runs of the experiment here suggested, that the direct photon is detected  significantly less frequently than expected from the \textit{a priori} probability $p_0$ given by \Eq{getthrough}, this may signal ``roughness'' of spacetime at Planck scale.  Without making specific hypotheses about quantum spacetime, we cannot estimate $\pi_*$.  However, it may be possible, by varying $n-1$, $L_1$ or $M$, to determine the critical scale above which the situation corresponds to a smooth spacetime.  This would then provide a check on Wheeler's conjecture as well as on the large extra dimensions idea.  Conversely, failure to detect any anomaly would not be inconsistent with the thermodynamic view of spacetime~\cite{Jacobson}.

\section{Calibration}\label{cal}

\begin{figure}
\includegraphics[width=3.4in]{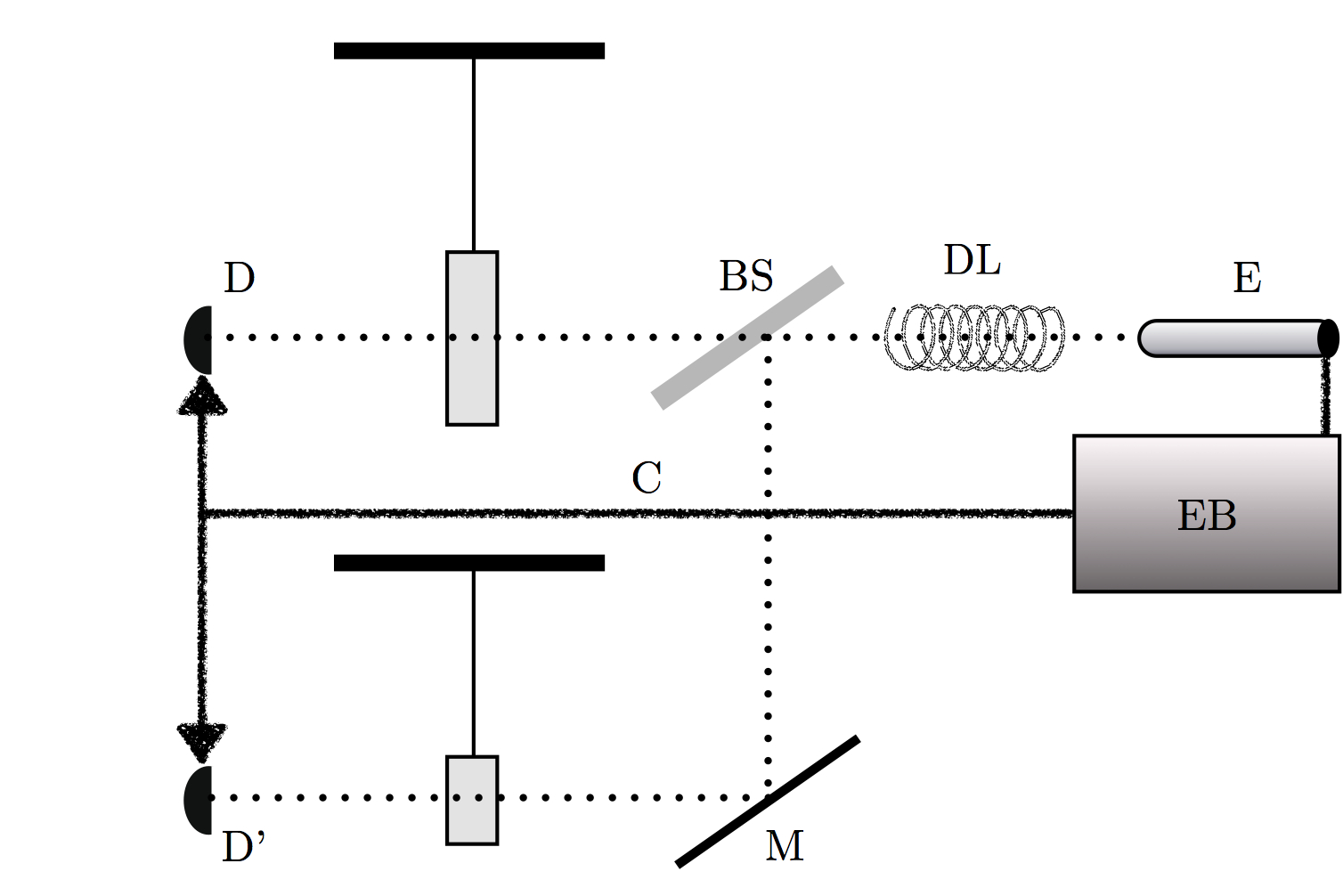}
\caption{\label{Fig2}Set up of suspended blocks showing (dotted) the alternative paths for the photon.  E is the single-photon emitter, D and D' are the single-photon detectors.  BS denotes the beamsplitter and M the mirror.  DL is the fiber optics delay line, and EB are the electronics that trigger D and D' through cable C.  The optical elements to widen the beam before and focus it after each block are left out for clarity.  In the real experiment the blocks would hang side by side.}
\end{figure}

Inaccuracies in the various optical parameters may confuse the probably small effect we are after.   One way to sidestep this hurdle is to supplement the basic setup of Fig.~\ref{Fig1} with a second, comparison, block of identical composition suspended side by side with the first, also by a thin fiber, and followed by its own single-photon detector D'.  For clarity the second block is displayed in the lower half of Fig.~\ref{Fig2}.  The dimensions of the second block should be adjusted so that the corresponding $\Delta X_0$ (for the same wavelength as earlier) is much longer than $\ell_P$.  For example, both blocks can be made of high lead glass with density $6\times 10^3\,{\rm Kg}\,{\rm m}^{-3}$ and $n=1.6$.  The first will have $L_1= 10^{-3}\,$ and $L_2= 5\times 10^{-3}\,$ so that $M=1.5\times 10^{-4}\, {\rm Kg}$ and $\Delta X_0= 1.98\times 10^{-35}\, $m.  The second block can have $L_1=L_2= 10^{-3}\,{\rm m}$ to which correspond $M=6\times 10^{-6}\, {\rm Kg}$ and $\Delta X_0= 4.96\times 10^{-34}\, {\rm m}\, \approx 30\, \ell_P$. 

The Newtonian force between the two example blocks set $0.2\,{\rm m}$ apart is $1.5\times 10^{-18}\, {\rm N}$.  Over the course of the photon transit this will impart either block a momentum $8\times 10^{-30}\, {\rm Kg\, m\, s}^{-1}$ which is a factor of
70 below the momentum acquired from the photon.  Thus mutual gravitation can be kept from being disruptive.

As shown in Fig.~\ref{Fig2}, the beam from the single photon emitter is directed through a 50-50 beam splitter and mirror assembly from which it continues to the blocks as shown in the figure.  One advantage of this configuration over the alternative in which a series of measurements is made using the first block followed by another series using the second, is that the effect of possible temporal drifts in the apparatus is minimized by working both arms of the setup essentially simultaneously.  

Now if a particular photon is ultimately detected by D, it has, with high probability (see Sec.~\ref{transits}), transited the upper block only once, and has caused a translation of it of order $\ell_P$, as per our example.  If instead the photon is detected by D', it has gone through the lower block, most probably without internal reflection, and has translated it by a distance some 30 times larger than $\ell_P$.  According to the ideas already mentioned, the first alternative is expected to be somewhat suppressed because of the non-smooth character of space-time at Planck scale.   Thus if the two arms of the setup in Fig.~\ref{Fig2} are perfectly balanced, events in which the photon is detected by D' should somewhat outnumber those in which it is detected by D. 

This asymmetry  might be small and might thus be swamped by unbalance of the beamsplitter-mirror assembly.  Relative calibration of the two-block setup is thus required.  This could be done by preceding the series of single-photon measurements with a separate experiment in which a macroscopic laser pulse is directed down the same paths as the single photons, and the relative intensities at D and D' are accurately measured (with a pair of more suitable counters which may be periodically interchanged to correct for differences in their efficiencies).  

While conclusions can be drawn statistically, following a series of single photon events, from the fraction of counts in D and D',  it may also prove interesting to study each event separately. To prevent confusion between a click in D or D' and an unrelated single photon emission from E, one can opt to trigger the detectors through cable C issuing from electronics EB which detect the emission.  This  signal can be allowed time to precede the photon by having the latter ``stored'' in a delay line, e.g. the coil of optical fiber DL in Fig.~\ref{Fig2}. 

\section{Sources of noise}  \label{which}

As in any experiment one must here contend with sources of noise which tend to cover up the phenomenon under investigation.
Confusion of the single photon in question with background light can be reduced to a tolerable level by proper shielding, cooling to suppress thermal optical  background, and the use of a narrowband filters centered on the single photon's wavelength placed at the detectors' inputs.  

We now consider noise of extra-terrestrial origin.  Cosmic ray hits on the blocks might have been a problem.  A cosmic ray carries such high momentum compared to the optical photon's that it would totally wash out the small displacement if they arrived together.  But the most abundant cosmic rays (protons with energies $> 0.4~{\rm GeV}$) account for a flux less than $10^4\,{\rm s}^{-1}\,{\rm m}^{-2}\,{\rm sr}^{-1}$~\cite{simpson}.  Thus the chance of such a cosmic ray overlapping with the occasional optical photon transit is a totally ignorable $4\times 10^{-9}$ assuming that the single photons arrive at a rate $10^2\,{\rm s}^{-1}$.

What about solar neutrinos?  Recently the Borexino collaboration~\cite{borexino} employing $2.8\times 10^5\,{\rm Kg}$ of liquid scintillator as detector (as compared with our $1.5\times 10^{-4}\,{\rm Kg}$ block), managed to detect but a few neutrinos per day.  These neutrinos from the pep reaction account for about 0.25\% of the totality of neutrinos emitted by the sun.   From this we \textit{crudely} estimate the probability of any type solar neutrino interacting with the block in coincidence with one of those $ 10^2\, {\rm s}^{-1}$ single photons  at $2\times 10^{-12}$. Solar neutrinos are evidently not a problem here.

Assessing the noise caused by the hypothetical dark matter particles is made particularly difficult by the great variety of types speculated upon.  We take an empirical tack.  About the only positive experimental evidence for dark matter has been provided by the decade long DAMA/LIBRA experiment in the Grand Sasso tunnel~\cite{DAMA}.  That experiment now uses $2.4\times 10^2\,{\rm Kg}$ of Na\,I scintillator as detector for the weakly interacting dark matter particles thought to be trapped in the potential well of our galaxy. It is sensitive to the 2-6 KeV energy range. 

The DAMA/LIBRA collaboration claims to have detected a correlation of the number of events detected with the yearly motion of the Earth about the Sun.  Largest count numbers are seen in early summer, when the Earth $30\, {\rm Km}\, {\rm s}^{-1}$ orbital velocity vector most nearly aligns with the solar $2\times 10^2\, {\rm Km}\, {\rm s}^{-1}$ velocity around the galactic center; lowest count rate comes 6 months later.  The instrument reveals a sinusoidal variation in between these dates with amplitude $1\times10^{-2}\, {\rm counts}/{\rm day}\, {\rm Kg}^{-1}\,{\rm KeV}^{-1}$.  Evidently the full mean event rate must be a factor $200/30$ larger.  

Making the exaggerated assumption that the spectrum of weakly interacting dark matter particles extends to $1\,$MeV without dropping at all makes the rate $66\, {\rm counts}/{\rm day}\, {\rm Kg}^{-1}$.  The probability that a DAMA/LIBRA type event will occur in our $1.5\times 10^{-4}\,{\rm Kg}$ block in coincidence with the crossing of one of those $ 10^2\, {\rm s}^{-1}$ single photons can thus be estimated as being below $5\times 10^{-17}$.  Even without delving into the other possible types of dark matter particles that are speculated upon, it should be clear that as sources of noise in the experiment contemplated here they are ignorable.

The most troublesome source of noise is thermal jitter of the primary block's c.m.  Let us begin with a na\" ive analysis of it.  We compare the r.m.s. speed of the c.m. in thermal equilibrium at temperature $T$, $V_t=\sqrt{3k_{\rm B} T/M}$, with the speed imparted to the c.m. by the momentum transfer $\Delta p$.  The two become equal at the crossover temperature $T_c$:
\be
k_{\rm B} T_c=\frac{(n-1)^2}{3n^2}\frac{\hbar\omega}{Mc^2}\, \hbar\omega\,,
\label{T}
\ee
and we obviously want to operate well below $T_c$.  
For the example of the first block in Sec.~\ref{cal} we already mentioned that $(\hbar\omega/Mc^2)\approx 3\times 10^{-32}$ so that $k_{\rm B} T_c\approx 4.34\times 10^{-33}\,$eV or $T_c\approx 5.04\times  10^{-29}\, {}^{\rm o}$K.  This is hopelessly  below foreseeable experimental reach!  Tweaking the parameters over a practical range does not help much.  However, the experiment is not thereby made unfeasible; the above analysis is simply improperly focused.

The thermal jitter of the c.m. is maintained by collisions of ambient gas atoms or molecules and thermal photons with the block.  The random block speed $V_t$  estimated above is acquired after so many collisions that thermodynamic equilibrium has been reached between gas, radiation and block. Let us first ignore the thermal photons and focus on impacts by atoms or molecules.  In between collisions the block's c.m. moves uniformly, and can be said to be at rest in \emph{some} Lorentz reference frame.  If the gas is very tenuous, collisions may be rare enough that, with high probability, the c.m. velocity acquired by the block from the optical photon does not get changed during the photon transit.  It then becomes irrelevant that there is an overall large thermal noise.  Put another way, the obstruction to the experiment is not the c.m. thermal motion \emph{per se}, but the frequency with which significant changes occur in it.

Let us introduce the length $L$ defined by
\be
L^2=2(2 L_1L_2+L_2{}^2)\,.
\ee
$L^2$ is just the total surface area of the block.  Now the number density of ambient molecules is $\varrho/m_*$,  where $\varrho$ is the mass density of the gas, while $m_*$ is the mass of an atom or  molecule.   Thus a good estimate of the number of hits per unit time of ambient molecules on the block is $(\varrho/m_*) \sqrt{3k_{\rm B} T/m_*}\,L^2$.   Let us require that the \textit{mean} number of hits over the transit time $n L_1/c$ is much smaller than unity.  We obtain the condition
\be
\Pi \equiv n L^2 L_1  P \sqrt{\frac{3}{m_* c^2 k_{\rm B}T}}\ll 1\ ,
\label{Pi}
\ee
where we replaced $\varrho/m_*$ by $ P/(k_{\rm B}T)$ in accordance with the ideal gas law, an excellent approximation at the low pressures $P$ we have in mind. 

We may obviously interpret $\Pi$ as the probability of a hit on the block during photon transit.   In those rare trials when the block is hit during photon transit by an atom or molecule, the momentum thus transferred to the block, of order $\sqrt{3m_* k_{\rm B}T}$, well exceeds that given by the photon (unless $T$ drops to near $10^{-6}\,{}^{\rm o}$K).  In such case the block receives a permanent momentum increment much larger than $\Delta p$  in \Eq{follow} and its translation increases constantly, so that eventually it is much bigger than Planck's length.  Consequently we do not expect any photon transmission anomaly in this case.  

By contrast, in the more frequent trials with no hit, the discussion in Sec.~\ref{cal} applies literally, and the probability of photon back reflection in the upper path of Fig.~\ref{Fig2} is expected to be enhanced. Thus when condition (\ref{Pi}) is satisfied, a series of single-photon trials will result in the above described asymmetry between the counts in detectors D and D'.  Can $\Pi$ be made small in practice?

Pressures down to $1.3\times 10^{-11}\,$Pa could be reached in low pressure labs already two decades ago~\cite{Ishi}, and $10^{-11}\,$Pa can routinely be obtained today with off the shelf equipment.  With $m_*=4$ a.m.u. (appropriate to Helium which, being inert, should be the gas of choice for the block's environment) and assuming room temperature  $T=300\,{}^{\rm o}$K  we find $\Pi=9\times 10^{-4}   $ for the first block and $\Pi=1\times 10^{-4}   $ for the second in the example of Sec.~\ref{cal}.  For various reasons it may be necessary to work at low temperatures.  It is germane here to note that pressures as low as $6.7\times 10^{-15}\,$Pa were indirectly measured in a $4\,{}^{\rm o}$K (liquid He temperature) vacuum system already 20 years ago~\cite{Gabrielse}.  But even at $P=10^{-11}\,$Pa with $T=4\,{}^{\rm o}$K we find $\Pi=1\times 10^{-2}$ for the first block and $\Pi=9\times 10^{-4}$ for the second.  It is thus experimentally feasible to reduce the atom hit probability during single photon transit to 1\% or less, which makes most single-photon events usable.

Let us now consider the effects of thermal photon noise.  The mean number density of thermal photons is~\cite{LLSP}
\be
N=0.244\left(\frac{k_B T}{\hbar c}\right)^3=2.03\times 10^7\, T({}^{\rm o}K)^3\,{\rm m}^{-3}\,.
\label{numb}
\ee
We estimate the number of thermal photon hits on the block \textit{per unit time}  as $N c L^2$.  For $T=300\,{}^{\rm o}$K ($T=4\,{}^{\rm o}$K) the first block in our example of Sec.~\ref{cal} receives $\sim 6\times 10^7$ (140) hits during the optical photon transit.

Now the photon number spectrum peaks at wavelength~\cite{LLSP}
\be
\lambda_{\rm peak}=\frac{1.60\, \hbar c}{k T}=\frac{0.00367\,{\rm m}}{T({}^{\rm o}K)}\ . 
\label{lambda}
\ee
For the range $4{}^{\rm o}{\rm K}$--$300{}^{\rm o}{\rm K}$ the photons are in the microwave to extreme infrared range.  Their frequencies lie well below most of the oscillator frequencies we could associate with the atomic constitutes of transparent dielectrics.  Hence the relevant value of $n$  is close to the zero frequency one, which is usually a few times unity.   Fresnel's formulae \Eq{Fre} thus predict that the photons will be most probably be reflected by the dielectric (actually scattered by the block).

Consequently, each thermal photon that hits the block imparts to it momentum in a \emph{random} direction of order of the momentum corresponding to $\lambda_{\rm peak}$, that is $\sim 3.92\,  k_{\rm B} T/c$.  Consequently, the total  momentum $\delta p$  so acquired from thermal photons \textit{during the optical photon's transit} will scale up with the \textit{square root} of the number of hits.  The final result is
\be
\delta p \approx 1.9\, \hbar\, (n L_1 L^2)^{1/2}\left(\frac{k_B T}{\hbar c}\right)^{5/2} .
\label{random}
\ee
 
We find for $T=300\,{}^{\rm o}$K that $\delta p=4.3\times 10^{-25} \ {\rm Kg}\, {\rm m}\, {\rm s}^{-1}$ and $1.3\times 10^{-25}\ {\rm Kg}\, {\rm m}\, {\rm s}^{-1}$ for the first and second blocks, respectively.  For $T=4\,{}^{\rm o}$K the mean momenta acquired are  instead $8.6\times 10^{-30} \ {\rm Kg}\, {\rm m}\, {\rm s}^{-1}$ and $2.5\times 10^{-30} \ {\rm Kg}\, {\rm m}\, {\rm s}^{-1}$, respectively.  By comparison (Sec.~\ref{ideal}) the optical photon's momentum transfer to either block is $\Delta p = 5.6\times 10^{-28}\ {\rm Kg}\, {\rm m}\, {\rm s}^{-1}$.  Obviously the experiment is rendered impossible at room temperature by thermal photon noise, but looks feasible at $T=4\,{}^{\rm o}$K for which that noise is only $1\%$ of the signal. 

One worry remains.  The momentum in \Eq{random}, unlike the optical photon's, is deposited permanently in the block.  When the optical photon exits, the block c.m. continues to move in its original rest frame , albeit much slower than during the transit.  Thus there is no exact sense in which the c.m. gets translated by a definite distance.  However, since $\delta p\ll \Delta p \approx \hbar\omega/c$ this may not be important.  After all, by placing the detectors D and D' near enough to the blocks one may bring the experiment to a close with the detection of each optical photon rather soon after its transit.  Thus the continuing slow c.m. drift may be irrelevant.

Any doubts of the kind just mentioned may be allayed by cooling the blocks and their environment further.  For $T\geq 4\,{}^{\rm o}$K the typical thermal photon wavelength, $\lambda_{\rm peak}$, is smaller than the larger dimension of the first block in our example.  It is then a good approximation to equate the optical crossection of the block and its geometric crossection\cite{Spitzer}.  Once the temperature is brought an order of magnitude below $ 4\,{}^{\rm o}$K, $\lambda_{\rm peak}\gg L_2$ and Rayleigh's regime sets in~\cite{LLED}, with the optical crossection scaling as $\lambda^{-4}$.  In view of \Eqs{numb}-(\ref{lambda}) the contemplated reduction of $T$ by an order of magnitude would cut the rate of thermal photon hits by a factor of $10^3\cdot10^4=10^7$.  As a consequence the probability of a thermal photon hit during photon transit would be reduced to $10^{-5}$.  Thus most of the single optical photon events would be free of noise, and the statistics of counts in D and D' would need no correction on account of block thermal noise.

\section{Summary}\label{sum}

The feasibility of translating the c.m. of a macroscopic block of dielectric by a distance of order Planck's length without first localizing it has been demonstrated.  This translation is not measured but inferred by momentum conservation involving a single optical photon which crosses the block. It is argued that such translation may occasionally be at odds with the non-smooth texture of spacetime on Planck scale.  Contradiction is then avoided if the photon is reflected by the block more often than predicted by classical electrodynamics.   An experimental set up  is proposed to detect this transmission anomaly, even if tiny; it compares the effect of the same photon on the above mentioned block and on a similar block which gets translated a distance much larger than Planck's.  It is shown that the thermal noise that might compromise the experiment is sufficiently suppressed by operating the blocks in an ultrahigh vacuum at temperature below $\sim 0.5 \,{}^{\rm o}$K.  

\acknowledgments

I thank participants of the conference ``Forty Years of Black Hole Thermodynamics'' in Jerusalem, particularly Raphael Bousso, Marek Karliner and Bill Unruh and those of the workshop ``Horizons of Quantum Physics'' in Taipei, especially Wei-Tou Ni, Lajos Djosi and Thomas Jennewein, as well as Alexander Feinstein, Igor Pikovski and Carlo Rovelli for their interest and illuminating criticism, comments and suggestions.  The present work is supported by grant 24/12 of the Israel Science Foundation.

  \appendix*
  
  \section{}

The discussion in Sec.~\ref{ideal} seems to ignore the incompatibility of measurements of momentum and position of the block.  However, a \textit{shift} in the block's c.m. coordinate can in fact be measured simultaneously with the block momentum.  

Let the operators for coordinates and kinematic momenta of particles in the block be $\hat {\bm r}_i$ and $\hat {\bm p}_i$.
With $M\equiv \sum_i m_i$, the center of mass coordinate is 
\be
\hat {\bm R}(t)=\frac{1}{M}\sum_i m_i\hat {\bm r}_i(t) 
\ee
whereas the total kinematic momentum is
\be
\hat{\bm P}(t)=\sum_i \hat {\bm p}_i(t).
\ee
It follows immediately from the canonical commutators $[\hat x_i,\hat p_{xj}]=\imath\hbar\delta_{ij}$, $[\hat x_i,\hat p_{yj}]=0$, etc. that  
\be
[\hat {\bm R}(t),\hat {\bm P}(t)]=\imath \hbar I\,,
\label{canonical}
\ee
 where $I$ is the $3\times 3$ unit tensor (dyad).

Let us designate a time $t_i$ as immediately following the ingress of the photon while $t_f$ is defined as a time immediately preceding its egress, so that the block's translation from time $t_i$ to $t_f$ is
\be
\Delta \hat {\bm X} \equiv \hat {\bm R}(t_f)-\hat {\bm R}(t_i).
\ee
We are interested in $[\Delta \hat {\bm X},\hat {\bm P}]$.  

Nonrelativistically the block's Hamiltonian may be written as
\be
\hat H =\sum_i  \half \frac{\hat{\bm p}_i{}(t)^2}{m_i}+\sum_{\rm a}\sum_{i\neq j}V_{\rm a}\big(\hat{\bm r}_i(t)-\hat{\bm r}_j(t)\big)
\ee
where $V_{1}, V_{2}, \cdots$ are the potentials for interactions of proton-proton Coulomb, proton-proton nuclear, neutron-neutron nuclear, proton-neutron nuclear, electron-proton Coulomb and electron-electron Coulomb types.  The nuclear interactions are approximated as being two-body and depending  on positions only through the vector distance between the particles (not necessarily leading to central forces).  The range of particles over which the summation over $i$ and $j$ takes place is, of course, different for each kind of potential. 

Now notice that for all $i$ and $j$
\bea
 \left[\hat{\bm P}(t),V\big(\hat{\bm r}_i(t)-\hat{\bm r}_j(t)\big)\right]= 
 \nonumber
 \\
 V'\cdot\left([\hat{\bm p}_i(t),\hat{\bm r}_i(t) ]-[\hat{\bm p}_j(t),\hat{\bm r}_j(t) ]\right)=0.
\eea
The photon Hamiltonian, one for a quasiparticle, will not depend on the $\hat{\bm p}_i{}(t)$ or $\hat{\bm r}_i{}(t)$.  Hence the full Hamiltonian commutes with $\hat{\bm P}(t)$, so that the total block momentum is conserved as long as the photon is inside it.

Let us now work out the Heisenberg equation 
$\imath\hbar d\hat {\bm R}/dt=[\hat{\bm R},\hat H].$
Obviously $\hat {\bm R}(t)$ commutes with the potentials $V_{\rm a}$.  From the canonical commutation relations we deduce in addition that
\be
\left[\hat{\bm R}(t), \half \frac{\hat{\bm p}_i{}(t)^2}{m_i}\right]=\frac{\imath\hbar I \hat{\bm p}_i{}(t)}{M}\Longrightarrow \left[\hat{\bm R}(t), \hat H\right]=\frac{\imath\hbar I \hat{\bm P}}{M}
\ee
Integration of the mentioned Heisenberg equation gives $\hat {\bm R}(t)=\hat {\bm R}(t_i)+(t-t_i)\,\hat {\bm P}/M$ for $t_i<t<t_j$.  It follows that $\Delta \hat {\bm X}$ is proportional to $\hat {\bm P}$, so that $\Delta \hat {\bm X}$ and $\hat {\bm P}$ commute.  Thus the relevant variables in the experiment are compatible, and we can simultaneously ascribe precise values to the block's momentum and to $\Delta X_0$.

\end{document}